\documentclass[
  aps,
  pra,
  reprint,
  superscriptaddress,
  longbibliography,
  floatfix,
  amsmath,
  amssymb
]{revtex4-2}

\usepackage{amsfonts}
\usepackage{amsthm}
\usepackage{mathtools}
\usepackage{mathrsfs}
\usepackage{bm}
\usepackage{bbm}

\usepackage{braket}
\usepackage{graphicx}
\usepackage{dcolumn}
\usepackage{booktabs}
\usepackage{array}
\usepackage{multirow}

\usepackage[dvipsnames]{xcolor}
\usepackage{tikz}
\usetikzlibrary{arrows.meta,bending}
\usepackage[
  colorlinks=true,
  linkcolor=blue,
  urlcolor=blue,
  citecolor=blue
]{hyperref}
\usepackage[nameinlink,noabbrev]{cleveref}

\newcommand{\workrange}{\mathcal{C}}

\newtheorem{theorem}{Theorem}

\newtheorem{lemma}{Lemma}

\begin{document}
\title{Harnessing Non-Markovian Memory for Quantum-Battery Capacity}

\author{Jun-Li Jiang}
\affiliation{School of Information Science and Technology, Southwest Jiaotong University, Chengdu 610031, China}

\author{Xin-Zhu Liu}
\affiliation{School of Information Science and Technology, Southwest Jiaotong University, Chengdu 610031, China}

\author{Xiu-Yong Ding}
\affiliation{School of Information Science and Technology, Southwest Jiaotong University, Chengdu 610031, China}

\author{Xue Yang}
\email{yx12290552@swjtu.edu.cn}
\affiliation{School of Information Science and Technology, Southwest Jiaotong University, Chengdu 610031, China}

\author{Ming-Xing Luo}
\email{mxluo@swjtu.edu.cn}
\affiliation{School of Information Science and Technology, Southwest Jiaotong University, Chengdu 610031, China}
\affiliation{Hefei National Laboratory, University of Science and Technology of China, Hefei, Anhui 230088, China}

\date{\today}

\begin{abstract}
Non-Markovianity is commonly characterized through
information-theoretic signatures of memory. Here we take a
thermodynamic perspective and ask whether such memory can be
operationally harnessed to enhance quantum-battery capacity.
Within the process-tensor framework, we compare memoryless and
memory-enhanced  strategies and define their performance
gap as the memory-assisted capacity advantage of a multitime process. We prove that this advantage vanishes for
Markovian process tensors and is bounded by a trace-distance measure of
process non-Markovianity. In a complementary continuous-time setting, we show using random
telegraph dephasing that memory accumulated during the evolution can
enhance capacity retention relative to a matched Markovian reference
under the same control class.  These results identify
non-Markovian memory as an operational resource for quantum-battery
capacity.

\end{abstract}

\maketitle

\section{Introduction}
\label{sec:introduction}

Open quantum systems inevitably interact with their environments, and
the resulting dynamics can retain information about the system's past \cite{RevModPhys.88.021002,RevModPhys.89.015001,10.1093/acprof:oso/9780199213900.001.0001}.
Such memory effects are commonly associated with quantum
non-Markovianity and have traditionally been characterized through
information backflow, divisibility properties, and related
information-theoretic criteria \cite{Rivas2014_RPP,Li2018,PhysRevLett.103.210401,PhysRevLett.112.120404,PhysRevLett.117.050403,PhysRevA.96.032109,PhysRevLett.109.233601,PhysRevA.101.012123,PhysRevLett.127.060501,Chen2022,PhysRevLett.105.050403,PhysRevLett.112.210402,PhysRevA.82.042103,abiuso2023characterizing}.
These approaches provide powerful tools for identifying and quantifying
memory, but a distinct question is whether the memory itself can be
operationally exploited to improve a physical task. This motivates a
shift from characterizing non-Markovian memory to understanding its
utility as a resource.

Quantum batteries provide a natural setting for addressing this
question \cite{RevModPhys.96.031001,Ferraro2026}. They are quantum systems designed to store and release energy,
whose performance can be influenced by coherence, correlations, and
collective control
\cite{Binder2015,Campaioli2017,Ferraro2018,
GarciaPintos2020,Arjmandi2022,Hu2026}.
Among the corresponding figures of merit, battery capacity quantifies
the full energy range accessible from a state within its unitary orbit
\cite{Yang2023}, and thus characterizes its energy-storage capability
beyond the instantaneous stored energy.
Non-Markovian environments have been shown to enhance charging, work
extraction, or energy retention in specific open-system settings
\cite{PhysRevA.104.032207,PhysRevA.109.012224,
PhysRevA.111.032416,Tiwari2025,Bylicka2016,Kamin2020,PhysRevA.102.052223,Xu2026}.
These results naturally raise a more operational question: can
environmental memory itself be harnessed to generate an additional
battery-capacity advantage?

To address this question, memory must be characterized at the level of
multitime interventions rather than solely through reduced dynamical
maps. When a system is subject to multiple interventions, information
retained in the environment can mediate the influence of earlier
operations on later outputs, giving rise to inter-step temporal
correlations \cite{Taranto2019,White2020,PhysRevLett.132.060402}. Such correlations are naturally captured by the
process-tensor framework
\cite{PhysRevLett.120.040405,pollock2018non,PRXQuantum.2.030201,Keeling2026}, which provides an operational
description of memory beyond that inferred from divisibility properties
of reduced dynamical maps alone \cite{PhysRevLett.123.040401}.
Recent work has shown that non-Markovian multitime processes can provide
an advantage for work extraction under thermal operations
\cite{zambon2025quantum}. Here we address a distinct storage task:
whether temporal correlations can enlarge the battery's accessible
unitary energy range beyond a locally optimized benchmark, after
explicitly accounting for the capacity supplied by external control.

\begin{figure}[!t]
\centering
\includegraphics[width=0.8\linewidth]{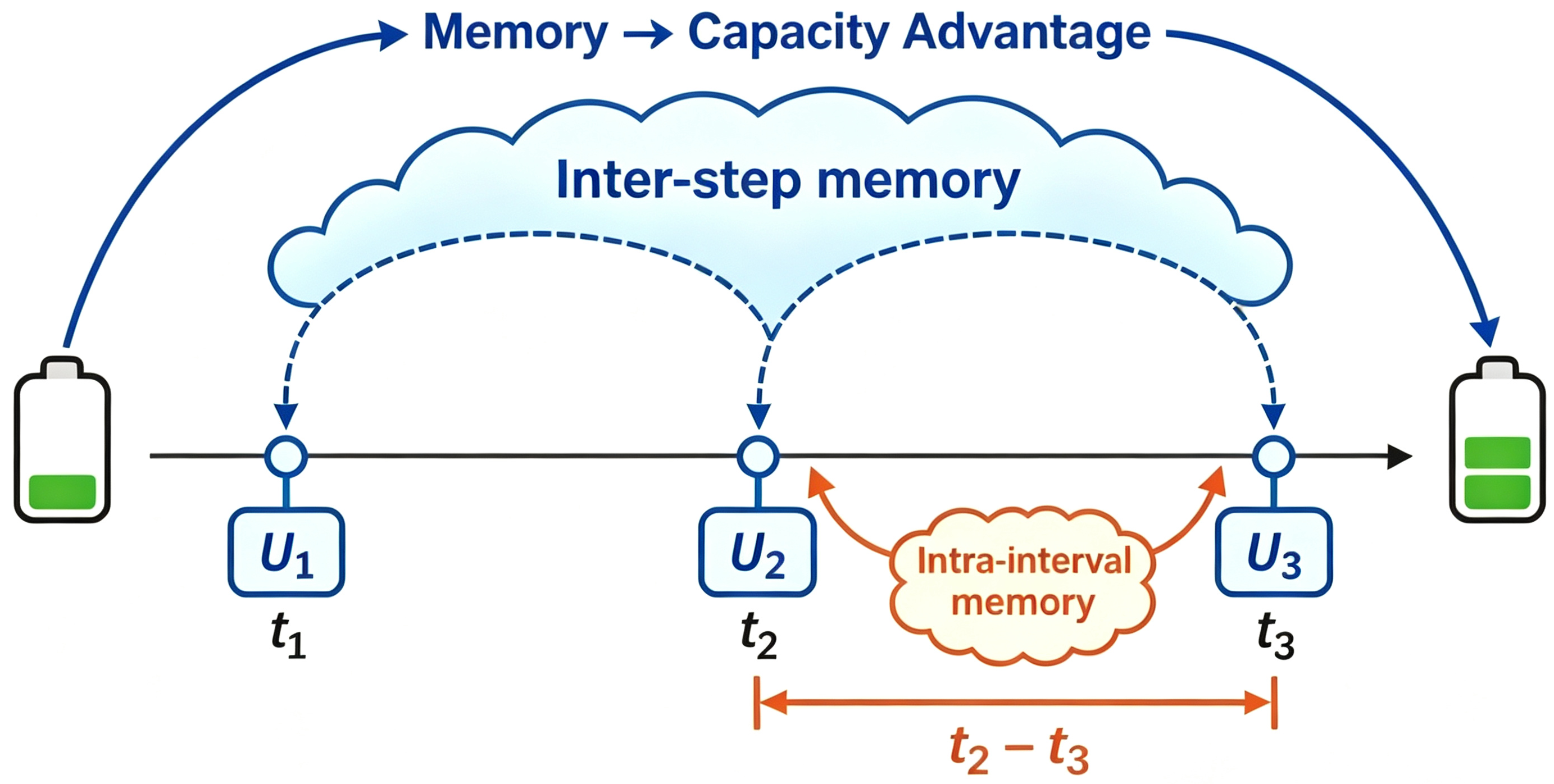}
\caption{Schematic illustration of memory-assisted quantum-battery capacity
enhancement in two complementary control settings: inter-step memory
across prescribed interventions and intra-interval memory accumulated
during continuous evolution.}
\label{Fig1}
\end{figure}

In this work, we investigate how non-Markovian memory can be
operationally converted into quantum-battery capacity advantages in two
complementary control settings, as illustrated in Fig.~\ref{Fig1}.
Within the process-tensor framework, memory is accessed across prescribed
intervention steps, which we refer to as inter-step memory. A locally
optimized protocol defines the memoryless benchmark, and its performance
gap from memory-enhanced control defines the memory-assisted capacity
advantage. We show that this advantage vanishes for Markovian process
tensors and is bounded by a distance-based measure of process
non-Markovianity; an explicit three-step qubit process demonstrates that
a strict advantage can occur. We then consider memory accumulated during
continuous evolution and accessed through an intermediate operation,
which we refer to as intra-interval memory. Using random telegraph
dephasing as a proof of principle, we show that such memory can enhance
capacity retention relative to a matched Markovian reference. Here,
inter-step and intra-interval specify where memory is accessed relative
to the chosen intervention structure, rather than distinct classes of
non-Markovianity.

The remainder of the paper is organized as follows.
Section~\ref{sec:framework} introduces the multitime capacity framework,
establishes the memory-assisted capacity advantage, and illustrates it
with a three-step process.
Section~\ref{secIII} considers the complementary continuous-time
realization using random telegraph dephasing.
Section~\ref{sec:IV} concludes with a discussion of the results and
their implications.

\section{Memory-assisted capacity advantage in multitime quantum processes} 
\label{sec:framework}

We now consider multitime quantum processes, where environmental memory
can mediate the influence of earlier interventions on later outputs.
Our aim is to determine whether such inter-step memory can be
operationally harnessed to generate a quantum-battery capacity advantage
and to quantify how this advantage is constrained by the memory of the
underlying process.

\subsection{Battery capacity and single-step capacity gain}
\label{subsec:capacity_gain}

We consider a quantum battery described by a state $\rho$ and a fixed
bare Hamiltonian $H$. The battery capacity \cite{Yang2023} quantifies the
total range of energies that can be accessed from $\rho$ through cyclic
unitary operations. It is defined as
\begin{align}
\workrange(\rho;H)
&=
\mathcal W(\rho;H)
+
\mathcal A(\rho;H)
\nonumber\\
&=
{\rm Tr}
\left[
(\rho^{\uparrow}-\rho^{\downarrow})H
\right],
\label{eq:capacity_def}
\end{align}
where $\mathcal W(\rho;H)$ and $\mathcal A(\rho;H)$ denote the ergotropy
and antiergotropy, respectively. Here $\rho^{\downarrow}$ and
$\rho^{\uparrow}$ are the passive and active rearrangements of $\rho$
with respect to the eigenbasis of $H$, corresponding to the minimum and
maximum energies accessible within the unitary orbit of $\rho$.
Unlike the instantaneous stored energy ${\rm Tr}(\rho H)$, the capacity
depends only on the spectrum of the state and the energy structure of
the Hamiltonian, and therefore characterizes the thermodynamic
potential available for future energy extraction or storage.

To quantify the capacity generated by a quantum process, we consider a
single-step channel $\Phi$ acting on a battery with Hamiltonian $H$. We
define its thermodynamic capacity gain as
\begin{align}
\Delta\workrange(\Phi;H)
:=
\max_{\rho}
\left\{
\workrange(\Phi(\rho);H)
-
\workrange(\rho;H)
\right\}.
\label{eq:channel_capacity_gain}
\end{align}
This quantity measures the maximal increase of the accessible energy
range generated by the channel $\Phi$.

\subsection{Process tensors and optimization strategies}
\label{subsec:process_tensor}

\begin{figure}[!t]
\centering
\includegraphics[width=0.8\linewidth]{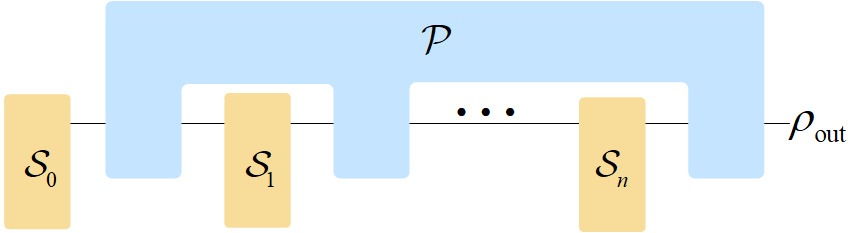}
\caption{Schematic representation of a fixed process tensor $\mathcal P$ with
prescribed intervention slots. A control strategy $\mathcal S$ specifies
the admissible operations inserted into these slots, producing the
accessible output state $\rho_{\rm out}=\mathcal P(\mathcal S)$.
Throughout the optimization, $\mathcal P$ is held fixed while
$\mathcal S$ is varied.}
\label{fig:comb}
\end{figure}

To formulate the multitime setting, we use the process-tensor framework
\cite{pollock2018non,PRXQuantum.2.030201}. As illustrated in
Fig.~\ref{fig:comb}, a process tensor $\mathcal P$ describes a
fixed uncontrolled multitime process with a prescribed sequence of
intervention slots. It encodes the operational influence of the underlying
system--environment dynamics across these slots, including any temporal
memory that can carry information from earlier interventions to later
outputs. A control strategy $\mathcal S$ specifies how the available slots
are used, including the preparation of input states and, depending on the
admissible control class, intermediate quantum operations and causal
feed-forward. Operationally, $\mathcal P$ acts as a higher-order map from
the chosen strategy to the corresponding accessible output state,
\begin{align}
\rho_{\rm out}=\mathcal P(\mathcal S).
\end{align}
For each optimization considered below, the underlying process
$\mathcal P$ is held fixed while the admissible strategy $\mathcal S$ is
varied. Our objective is therefore to determine whether different ways of
controlling the same multitime process can convert its temporal memory
into an additional quantum-battery capacity gain.

A process is Markovian when every causal break completely
removes the influence of earlier interventions, such that the future
dynamics depends only on the state re-prepared at the break. Any
remaining dependence of future outputs on previous interventions
indicates the presence of temporal memory
\cite{PhysRevLett.120.040405,pollock2018non}. This multitime description captures operational history dependence that
is not accessible from divisibility properties of reduced dynamical maps
alone \cite{PhysRevLett.123.040401}.

Throughout this section, all process input and output systems are taken
to be isomorphic copies of the same finite-dimensional battery with
identical local Hamiltonian $H$, so that their capacities are evaluated
with respect to a common energetic reference. To isolate the capacity
generated by the process from that supplied by external control, the
single-step gain $\Delta\mathcal C(\Phi;H)$ is also used to quantify the
capacity cost of each controller-implemented map $\Phi$. The net gains
defined below therefore subtract the capacity supplied by the controller.

\subsubsection{Memoryless sequential optimization}

To isolate the capacity advantage attributable to temporal memory, we
first introduce a sequential benchmark, which we denote the
\emph{memoryless sequential} (ML) strategy. We regard
each intervention slot as a separate optimization step. Once the
preceding input trajectory $\bm r_{<i}^{\rm ML}$ has been fixed, the
process induces a conditional mapping
$\mathcal P_{i|\bm r_{<i}^{\rm ML}}$ at step $i$. Following the
single-step capacity gain defined in Eq.~\eqref{eq:channel_capacity_gain},
the controller freshly prepares the input at this step to maximize the
immediate capacity gain,
\begin{align}
\rho_i^{\rm ML}
\in
\arg\max_{\rho}
\Big[
\workrange\!\left(
\mathcal P_{i|\bm r_{<i}^{\rm ML}}(\rho);H
\right)
-
\workrange(\rho;H)
\Big].
\label{eq:SM_ML_input}
\end{align}
Here,
$\bm r_{<i}^{\rm ML}
=(\rho_1^{\rm ML},\ldots,\rho_{i-1}^{\rm ML})$
denotes the inputs selected at the preceding steps. Thus, the choice at
step $i$ is locally optimal: it maximizes the current capacity gain
without accounting for how this choice may affect the conditional
mappings encountered at later steps.

The total gain obtained from this optimization is
\begin{align}
\Delta\workrange^{\rm ML}(\mathcal P;H)
&=
\sum_{i=1}^{n}
\Big[
\workrange\!\left(
\mathcal P_{i|\bm r_{<i}^{\rm ML}}
(\rho_i^{\rm ML});H
\right)
-
\workrange(\rho_i^{\rm ML};H)
\Big]
\nonumber\\
&=
\sum_{i=1}^{n}
\Delta\workrange\!\left(
\mathcal P_{i|\bm r_{<i}^{\rm ML}};H
\right).
\label{eq:SM_ML_gain}
\end{align}
Here, ``memoryless'' refers only to the
absence of cross-time coordination in the control choices: the
conditional mapping at each step may still depend on earlier inputs and
therefore retain temporal memory of the underlying process
$\mathcal P$.

\subsubsection{Memory-enhanced optimization}

Having established the memoryless sequential strategy as a benchmark,
we now ask whether, for the same fixed process $\mathcal P$, coordinating
control choices across different intervention slots can convert
inter-step temporal memory into an additional battery-capacity gain.  We refer to strategies
that exploit such cross-time coordination as \emph{memory-enhanced} (ME)
strategies. Below, we consider three increasingly general strategies,
obtained by progressively relaxing the restrictions of the memoryless
sequential
optimization.

We first consider the simplest extension of the memoryless
sequential strategy, which we
refer to as \emph{joint optimization} (J). Rather than choosing each input
greedily according to its immediate capacity gain, we treat the full
input trajectory
\begin{align}
\bm r=(\rho_1,\ldots,\rho_n)
\end{align}
as a single optimization variable and select it at the outset so as to
maximize the sum of the local capacity gains over all intervention
steps. The output generated at each step is still evaluated locally;
thus, the joint strategy changes only how the inputs are coordinated
across time. The corresponding gain is
\begin{align}
\Delta\workrange^{\rm ME}_{\rm J}(\mathcal P;H)
=
\max_{\bm r}
\sum_{i=1}^{n}
\Big[
\workrange\!\left(
\mathcal P_{i|\bm r_{<i}}(\rho_i);H
\right)
-
\workrange(\rho_i;H)
\Big].
\label{eq:SM_MEJ}
\end{align}
This is the simplest setting in which an early input need not maximize
its immediate capacity gain, but may instead be chosen for its effect
on the conditional mappings encountered at later steps, thereby
increasing the total gain over the full trajectory. Since the ML input
trajectory is itself an admissible choice in the joint optimization,
\begin{align}
\Delta\workrange^{\rm ME}_{\rm J}(\mathcal P;H)
\ge
\Delta\workrange^{\rm ML}(\mathcal P;H).
\label{eq:MEJ_ge_ML}
\end{align}

We next extend the joint optimization strategy by allowing the outputs generated at
different steps to be retained and read out collectively. The motivation
is to determine whether correlations among the retained outputs of the
same multitime process can provide an additional capacity advantage
beyond the local readout used in the joint strategy. We refer to this
extension as \emph{global optimization} (G). The full input trajectory
\begin{align}
\bm r=(\rho_1,\ldots,\rho_n)
\end{align}
is still chosen non-adaptively before the process begins, as in the
joint strategy, but all output systems are stored until the final time.
The protocol may then choose between local readout and a collective
readout of the retained outputs.

Let
$\sigma=\mathcal P(\mathcal S_{\bm r})$
denote the joint state of all output systems retained for the final
readout, and let
$\sigma_i={\rm Tr}_{\bar i}\sigma$
be its marginal on the $i$-th output. For a collective readout, we use
the additive Hamiltonian
\begin{align}
H_{\rm out}
=
\sum_{i=1}^{n}H^{(i)},
\qquad
H^{(i)}\equiv H,
\end{align}
where $H^{(i)}$ acts on the $i$-th output system.
A collective readout, however, can increase the battery capacity even
for the uncorrelated product state with the same local marginals.
Therefore, to avoid attributing this purely global spectral-rearrangement
effect to temporal correlations, we define
\begin{align}
\Delta_{\rm prod}^{\workrange}(\sigma;H_{\rm out})
:=
\workrange\!\left(
\bigotimes_{i=1}^{n}\sigma_i;H_{\rm out}
\right)
-
\sum_{i=1}^{n}
\workrange(\sigma_i;H^{(i)}).
\label{eq:SM_Delta_prod}
\end{align}
The corresponding correlation-adjusted collective readout value is
\begin{align}
&\workrange(\sigma;H_{\rm out})
-
\Delta_{\rm prod}^{\workrange}(\sigma;H_{\rm out})
\nonumber\\
&=
\sum_{i=1}^{n}\workrange(\sigma_i;H^{(i)})
+
\workrange(\sigma;H_{\rm out})
-
\workrange\!\left(
\bigotimes_{i=1}^{n}\sigma_i;H_{\rm out}
\right).
\label{eq:SM_adjusted_collective}
\end{align}
Since local readout should remain an admissible option, we define the
effective readout capacity as
\begin{align}
&\workrange_{\rm read}(\sigma;H_{\rm out})
:=
\sum_{i=1}^{n}
\workrange(\sigma_i;H^{(i)})
\nonumber\\
&+
\max\Bigg\{
0,\,
\workrange(\sigma;H_{\rm out})
-
\workrange\!\left(
\bigotimes_{i=1}^{n}\sigma_i;
H_{\rm out}
\right)
\Bigg\}.
\label{eq:SM_Cread}
\end{align}
Thus the collective readout contributes only when the actual multi-output state has a larger capacity than the product state with the same marginals.

The globally optimized memory-enhanced gain is then
\begin{align}
 \Delta\workrange_{\rm G}^{\rm ME}(\mathcal P;H)=\max_{\mathcal S_{\bm r}\in\mathbb S_n^{\rm G}} \bigl(\workrange_{\rm read} (\mathcal P(\mathcal S_{\bm r});H_{\rm out})-\sum_{i=1}^{n}\workrange(\rho_i;H) \bigr). \label{eq:SM_MEG} 
\end{align} 
Here, $\mathbb S_n^{\rm G}$ denotes the set of non-adaptive strategies
specified by a preselected input trajectory together with passive storage
of all outputs. Since local readout is included as an admissible option,
restricting this strategy class to local readout recovers the joint
optimization protocol.

We finally ask whether the control can be made adaptive across
successive intervention steps. In both the joint and global strategies,
the complete input trajectory is fixed before the process begins.
However, if an earlier process output carries information about the
temporal memory of $\mathcal P$, it may be advantageous to use that
output to determine the input applied at the next step. We therefore
extend the global strategy to a restricted \emph{quantum-comb optimization}
(C), in which causal feed-forward between successive process slots is
allowed.

To ensure that any resulting advantage is attributed to the temporal
memory of the process rather than to an independent memory carried by
the controller, we restrict the admissible combs to a class
$\mathbb S_n^{\rm C}$ with explicit controller-cost accounting. For any strategy in $\mathbb S_n^{\rm C}$, the output of each process
slot is assigned one of two roles. It may be passively retained for the final readout, in which
case it is not acted on by subsequent controls. Alternatively, it may
be processed by a single-system control map
$\mathcal E_i:S\to S$ to prepare the input to the next process slot,
in which case it is not retained for the final readout. The input
and output of $\mathcal E_i$ are associated with the same local
Hamiltonian $H$, and the capacity that can be supplied by this control
map is explicitly included in the controller cost through
$\Delta\workrange(\mathcal E_i;H)$.

An individual map $\mathcal E_i$ may employ ancillary degrees of freedom
in its physical realization, but these ancillas are discarded at the end
of that control step and cannot serve as persistent memory across
different intervention steps. We further require that the input entering
each process slot be uncorrelated with all registers already retained for
the final readout. If $R_{i-1}$ denotes the collection of retained
registers before the $i$-th slot, then
\begin{align}
\Omega_{R_{i-1}S_i}
=
\Omega_{R_{i-1}}\otimes\rho_i .
\label{eq:SM_comb_factorization}
\end{align}
Thus, retained outputs act only as passive records, while adaptivity is
implemented solely through the charged single-system feed-forward
maps. The controller therefore contains no additional persistent memory whose
capacity contribution is left unaccounted for.

For a strategy $\mathcal S\in\mathbb S_n^{\rm C}$, let
$\mathcal P(\mathcal S)$ denote the final accessible state on the retained
output registers $B_1,\ldots,B_{m(\mathcal S)}$, with
$H_{\rm out}^{\mathcal S}
=
\sum_{j=1}^{m(\mathcal S)} H_{B_j}$.
Using the same correlation-adjusted readout rule introduced for the
global strategy, the comb-optimized memory-enhanced gain is defined as
\begin{align}
\Delta\workrange_{\rm C}^{\rm ME}(\mathcal P;H)
=
\max_{\mathcal S\in\mathbb S_n^{\rm C}}
\left[
\workrange_{\rm read}
\bigl(\mathcal P(\mathcal S);H_{\rm out}^{\mathcal S}\bigr)
-
\Delta\workrange(\mathcal S;H)
\right].
\label{eq:SM_MEC}
\end{align}
Here, $\Delta\workrange(\mathcal S;H)$ denotes the total capacity cost of
the controller. If
$\{\mathcal E_0,\mathcal E_1,\ldots,\mathcal E_{n-1}\}$
are the charged control maps of $\mathcal S$, with $\mathcal E_0$
preparing the first input, we define
\begin{align}
\Delta\workrange(\mathcal S;H)
:=
\sum_{i=0}^{n-1}
\Delta\workrange(\mathcal E_i;H).
\label{eq:SM_comb_cost}
\end{align}
When an output is retained rather than used for feed-forward, the next
input is prepared independently by a replacement map. For
$\mathcal E(\cdot)=\rho$, this convention gives
$\Delta\workrange(\mathcal E;H)=\workrange(\rho;H)$,
which recovers the input-cost accounting used in the joint and global
strategies.

By construction, the three memory-enhanced strategy classes are nested,
yielding
\begin{equation}
\Delta\workrange_{\rm J}^{\rm ME}(\mathcal P;H)
\leq
\Delta\workrange_{\rm G}^{\rm ME}(\mathcal P;H)
\leq
\Delta\workrange_{\rm C}^{\rm ME}(\mathcal P;H).
\label{eq:SM_ME_hierarchy}
\end{equation}
We therefore define the overall memory-enhanced gain by the most general
admissible class considered here,
\begin{equation}
\Delta\workrange^{\rm ME}(\mathcal P;H)
:=
\Delta\workrange_{\rm C}^{\rm ME}(\mathcal P;H).
\label{eq:SM_ME_def}
\end{equation}

\subsection{Genuine temporal-memory capacity advantage}
\label{subsec:memory_advantage}

The memory-enhanced and memoryless strategies defined above provide two
different levels of access to the same multitime process. To quantify
the thermodynamic value of temporal correlations, we define the
memory-assisted capacity advantage as
\begin{align}
\mathcal A_{\workrange}(\mathcal P;H)
:=
\Delta\workrange^{\rm ME}(\mathcal P;H)
-
\Delta\workrange^{\rm ML}(\mathcal P;H).
\label{eq:memory_advantage}
\end{align}
This quantity measures the additional battery capacity that can be
generated by exploiting temporal memory beyond a strategy that optimizes
each step locally. Importantly, the capacity supplied by external
control operations has already been removed through the capacity-cost
accounting introduced above. Therefore,
$\mathcal A_{\workrange}$ characterizes the net thermodynamic
contribution of temporal correlations in the process.

To relate this capacity advantage to the non-Markovianity of the
underlying multitime process, we use the distance-based measure
\cite{Berk2021resourcetheoriesof,PhysRevA.110.042210}
\begin{align}
N(\mathcal P)
:=
\frac{1}{2}
\min_{\mathcal Q\in\mathbb P_n^{\rm M}}
\sup_{\mathcal S\in\mathbb S_n}
\left\|
\mathcal P(\mathcal S)
-
\mathcal Q(\mathcal S)
\right\|_1 ,
\label{eq:process_NM_measure}
\end{align}
where $\mathbb P_n^{\rm M}$ denotes the set of Markovian $n$-step
process tensors and $\mathbb S_n$ the full set of compatible $n$-step
strategy combs. This quantity measures the minimal operational distance
between $\mathcal P$ and the set of Markovian processes.

The relation between temporal memory and capacity advantage is
summarized by the following theorem.

\begin{theorem}
For any Markovian $n$-step process tensor $\mathcal P_{\rm M}$,
\begin{align}
\Delta\workrange^{\rm ME}(\mathcal P_{\rm M};H)
=
\Delta\workrange^{\rm ML}(\mathcal P_{\rm M};H),
\label{eqn-Markovian-collapse}
\end{align}
equivalently,
$\mathcal A_{\workrange}(\mathcal P_{\rm M};H)=0$.
For an arbitrary $n$-step process tensor $\mathcal P$,
\begin{align}
0
\le
\mathcal A_{\workrange}(\mathcal P;H)
\le
2n(n+3)\Lambda N(\mathcal P),
\label{eqn-ME-ML-bound}
\end{align}
where $\Lambda$ is the spectral width of the local Hamiltonian $H$.
\end{theorem}

The proof is provided in Appendix~\ref{appendix:Theorem1}. For a
Markovian process, the memory-enhanced and memoryless strategies yield
the same capacity gain. Hence, any positive memory-assisted advantage
certifies that inter-step temporal memory has been operationally
converted into additional battery capacity. The upper bound further
constrains the achievable advantage by the non-Markovianity of the
underlying process.

\subsection{Three-step realization of a memory-assisted capacity advantage}

To make the mechanism of the memory-assisted capacity advantage
explicit, we consider a three-step qubit process in which the choice of
the first input involves a tradeoff between the immediate capacity gain
and the capacity accessible at later outputs through temporal memory.
This simple construction allows the memoryless and memory-enhanced
strategies introduced above to be compared directly.

\begin{figure}[!t]
\centering
\includegraphics[width=0.85\linewidth]{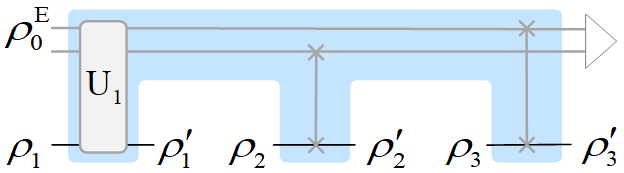}
\caption{
Fixed uncontrolled process $\mathcal P$ for the three-step construction.
The memory qubits $E_1$ and $E_2$ are initially maximally mixed and
interact with the system through $U_1$, followed by
$U_2={\rm SWAP}_{S_2E_1}$ and
$U_3={\rm SWAP}_{S_3E_2}$. Strategy insertion slots are omitted.
}
\label{fig:pt_hierarchy}
\end{figure}

The process $\mathcal P$, illustrated in
Fig.~\ref{fig:pt_hierarchy}, consists of a system qubit interacting
sequentially with two environmental memory qubits $E_1$ and $E_2$.
The environment is initially prepared in the maximally mixed state
\begin{align}
\rho_0^E
=
\frac{\mathbbm 1_{E_1E_2}}{4},
\end{align}
and all system and environmental qubits are assigned the same local
Hamiltonian
\begin{align}
H=E|1\rangle\langle1|.
\end{align}
The maximally mixed initial environment carries no battery capacity,
thereby excluding pre-existing environmental capacity as the origin of
the separation between the strategies.

The first interaction transfers information about the initial system
state into the environmental memories, which are subsequently read out
at the second and third steps. Specifically, at the first step the
system interacts with $E_1E_2$ through a unitary $U_1$. Let
$c=\sqrt3/2,
s=1/2 $.
For $j=00,01,10,11$, define
\begin{align}
U_1|0,j\rangle=|v_j\rangle,
\qquad
U_1|1,j\rangle=|w_j\rangle,
\end{align}
where $(|v_{00}\rangle,\cdots,|v_{11}\rangle)=(|000\rangle, |001\rangle, 
|010\rangle, c|100\rangle+s|111\rangle)$ and $(|w_{00}\rangle,\cdots,|w_{11}\rangle)=(|011\rangle, 
|101\rangle, |110\rangle$, $-s|100\rangle+c|111\rangle)$. 
The eight output vectors are orthonormal, so these relations define a
unitary $U_1$.

At the second and third steps, the information stored in the two memory
qubits is transferred to the corresponding system outputs through
\begin{align}
U_2={\rm SWAP}_{S_2E_1},
\qquad
U_3={\rm SWAP}_{S_3E_2}.
\end{align}
The fresh inputs at these two steps are swapped into registers that do
not influence the subsequent outputs. Since their capacities still
enter as input costs, their optimal choices are
\begin{align}
\rho_2=\rho_3=\frac{\mathbbm 1}{2}.
\end{align}

It therefore remains to optimize the first input. We write
\begin{align}
\rho_1
=
\frac12
\left(
\mathbbm 1+\vec r\cdot\vec\sigma
\right),
\qquad
\vec r=(x,y,z).
\end{align}
For the qubit Hamiltonian $H=E|1\rangle\langle1|$, its capacity is
\begin{align}
\workrange(\rho_1;H)
=
E\|\vec r\|.
\end{align}
A direct calculation gives the capacities of the three output marginals,
\begin{align}
\workrange(\rho_1';H)
&=
\frac{E}{2}\|\vec r\|,
\\
\workrange(\rho_2';H)
=
\workrange(\rho_3';H)
&=
\frac{E}{8}
\left|3z-\sqrt3\,x\right|.
\end{align}
These expressions expose the mechanism behind the memory-assisted
advantage: the first input affects not only the immediate output but
also the capacity later recovered from the two environmental memories.
The different optimization strategies introduced above therefore make
different choices for $\rho_1$, as we now show explicitly.

We first apply the ML strategy. Since it optimizes the first step
according to the immediate capacity gain,
\begin{align}
\workrange(\rho_1';H)-\workrange(\rho_1;H)
=
-\frac{E}{2}\|\vec r\|
\le 0,
\end{align}
the optimal choice is
$\rho_1^{\rm ML}=\mathbbm 1/2$. For this input, the state after
$U_1$ is $\mathbbm 1_{SE_1E_2}/8$, so both later memory readouts are
also maximally mixed. Hence
\begin{align}
\Delta\workrange^{\rm ML}(\mathcal P;H)=0.
\label{eq:app_ML_zero}
\end{align}

We next allow the full input trajectory to be optimized jointly.
Since the second and third inputs are optimally maximally mixed, the
joint gain depends only on the first-input Bloch vector,
\begin{align}
G_{\rm J}(\vec r)
&=
\workrange(\rho_1';H)
+\workrange(\rho_2';H)
+\workrange(\rho_3';H)
-\workrange(\rho_1;H)
\nonumber\\
&=
-\frac{E}{2}\|\vec r\|
+\frac{E}{4}|3z-\sqrt3 x|.
\end{align}
Using
\begin{align}
|3z-\sqrt3 x|
\le 2\sqrt3\,\|\vec r\|,
\end{align}
we obtain
\begin{align}
G_{\rm J}(\vec r)
\le
\frac{\sqrt3-1}{2}E\|\vec r\|
\le
\frac{\sqrt3-1}{2}E.
\end{align}
The bound is attained for
$\vec r_\ast=(-1/2,0,\sqrt3/2)$, giving
\begin{align}
\Delta\workrange_{\rm J}^{\rm ME}(\mathcal P;H)
=
\frac{\sqrt3-1}{2}E.
\label{eq:app_joint_gain}
\end{align}

We then include global storage of the three outputs and allow the
collective readout introduced above. With
$H_{\rm out}=H^{(1)}+H^{(2)}+H^{(3)}$, the three-output state has two
fourfold-degenerate eigenvalues
$(1+\|\vec r\|)/8$ and $(1-\|\vec r\|)/8$, while $H_{\rm out}$ has
degeneracies $1,3,3,1$. The rearrangement formula therefore gives
\begin{align}
\workrange(\rho_{123}';H_{\rm out})
=
\frac{3}{2}E\|\vec r\|.
\label{eq:app_global_capacity}
\end{align}
The corresponding net gain is
\begin{align}
G_{\rm G}(\vec r)
=
\workrange_{\rm read}(\rho_{123}';H_{\rm out})
-\workrange(\rho_1;H).
\end{align}
The local-readout branch satisfies
\begin{align}
\sum_{i=1}^{3}\workrange(\rho_i';H)
=
\frac{E}{2}\|\vec r\|
+\frac{E}{4}|3z-\sqrt3 x|
\le
\frac{3}{2}E\|\vec r\|,
\end{align}
and the correlation-adjusted collective branch is bounded by the same
global capacity. Hence
\begin{align}
G_{\rm G}(\vec r)
\le
\frac{3}{2}E\|\vec r\|
-E\|\vec r\|
\le
\frac{E}{2}.
\end{align}
For $\vec r_\ast=(-1/2,0,\sqrt3/2)$,
$\Delta_{\rm prod}^{\workrange}(\rho_{123}';H_{\rm out})=0$, and the
collective branch attains the bound. Therefore
\begin{align}
\Delta\workrange_{\rm G}^{\rm ME}(\mathcal P;H)
=
\frac{E}{2}.
\end{align}

Finally, consider the restricted comb strategy. Since global storage is
an admissible special case,
\begin{align}
\Delta\workrange_{\rm C}^{\rm ME}(\mathcal P;H)
\ge
\Delta\workrange_{\rm G}^{\rm ME}(\mathcal P;H).
\end{align}
For the present process, however, the second and third fresh inputs are
swapped into registers that do not influence the retained memory
outputs, so adaptive choices of these inputs cannot increase the
process-generated output capacity. Moreover, for any single-system
control map $\Phi:S\to S$,
\begin{align}
\workrange(\Phi(\sigma);H)
-
\Delta\workrange(\Phi;H)
\le
\workrange(\sigma;H),
\end{align}
so after subtracting the assigned controller cost, no control operation
can provide a positive net capacity contribution. It follows that the
restricted comb cannot outperform global storage in this example, and
therefore
\begin{align}
\Delta\workrange_{\rm C}^{\rm ME}(\mathcal P;H)
=
\Delta\workrange_{\rm G}^{\rm ME}(\mathcal P;H)
=
\frac{E}{2}.
\label{eq:app_comb_gain}
\end{align}

Combining the above results,
\begin{align}
\Delta\workrange^{\rm ME}
&=
\Delta\workrange_{\rm C}^{\rm ME}
=
\Delta\workrange_{\rm G}^{\rm ME}
=
\frac{E}{2}
\nonumber\\
&>
\Delta\workrange_{\rm J}^{\rm ME}
=
\frac{\sqrt3-1}{2}E
>
\Delta\workrange^{\rm ML}
=
0.
\end{align}
The example therefore makes the memory-assisted advantage explicit:
the ML strategy misses the benefit of sacrificing the immediate
first-step gain for a larger later return, whereas the coordinated
strategies exploit the temporal memory stored in the process.

\section{Memory-assisted capacity retention in continuous-time dynamics}
\label{secIII}

The preceding section treated inter-step memory across prescribed
intervention slots. Here we examine the complementary intra-interval
setting, where memory accumulated during continuous evolution is
accessed by a single intermediate control operation. As a proof of
principle, we consider a qubit under random telegraph dephasing and
compare its optimized retained capacity with that of a matched
Markovian reference under the same control family.

\subsection{Single-pulse capacity retention under RTN dephasing}
\label{subsec:RTN_retention}

We consider a qubit battery subject to random telegraph dephasing noise.
The initial state is chosen on the equator of the Bloch sphere,
\begin{align}
\ket{\psi_0}
=
\frac{\ket{0}+e^{{\rm i}\phi}\ket{1}}{\sqrt{2}},
\end{align}
and the stochastic Hamiltonian is
\begin{align}
H(t)
=
\varepsilon\sigma_z+\nu c(t)\sigma_z,
\label{eq:RTN_H}
\end{align}
where $c(t)=\pm1$ switches with rate $\xi$ and satisfies
$\langle c(t)c(t_0)\rangle
=
e^{-2\xi|t-t_0|}$.

Since the Hamiltonian is diagonal in the $\sigma_z$ basis, the
populations remain unchanged and the noise affects only the coherence.
Introducing
\begin{align}
G(t)
=
\left\langle
e^{2{\rm i}\nu\int_0^t c(s)\,ds}
\right\rangle_c ,
\label{eq:RTN_G}
\end{align}
the reduced state is
\begin{align}
\rho(t)
=
\frac12
\begin{pmatrix}
1 &
e^{-{\rm i}(2\varepsilon t+\phi)}G^*(t)
\\
e^{{\rm i}(2\varepsilon t+\phi)}G(t)
&
1
\end{pmatrix}.
\label{eq:RTN_rho}
\end{align}
For the bare Hamiltonian $H_0=\varepsilon\sigma_z$ \cite{Hadipour2024}, the corresponding
battery capacity is
\begin{align}
\workrange_{\rm free}(\rho(t);H_0)
=
2\varepsilon |G(t)|.
\label{eq:RTN_free_capacity}
\end{align}

Here, Markovianity refers to CP divisibility of the reduced dynamics.
Within this convention, the RTN model exhibits Markovian and
non-Markovian dephasing regimes. For $\xi>2\nu$, the dynamics is
Markovian (M), with
\begin{align}
G_{\rm M}(t)
=
e^{-\xi t}
\left[
\cosh(\Omega t)
+
\frac{\xi}{\Omega}\sinh(\Omega t)
\right],
\Omega=\sqrt{\xi^2-4\nu^2}.
\label{eq:RTN_G_M}
\end{align}
For $\xi<2\nu$, the dynamics is non-Markovian (NM), with
\begin{align}
G_{\rm NM}(t)
=
e^{-\xi t}
\left[
\cos(\delta t)
+
\frac{\xi}{\delta}\sin(\delta t)
\right],
\delta=\sqrt{4\nu^2-\xi^2}.
\label{eq:RTN_G_NM}
\end{align}
At the transition point $\xi=2\nu$, the two expressions coincide
continuously, yielding
\begin{align}
G(t)=e^{-\xi t}(1+\xi t).
\label{eq:RTN_G_critical}
\end{align}
Although the non-Markovian regime exhibits oscillations and temporary
revivals of $|G(t)|$, in both regimes
\begin{align}
\lim_{t\to\infty}G(t)=0,
\qquad
\lim_{t\to\infty}\workrange_{\rm free}(\rho(t);H_0)=0.
\label{eq:RTN_free_asymptotic}
\end{align}

Thus, the transient revivals generated by non-Markovianity do not by
themselves yield a capacity that survives at long times. To make such
transient dynamical features operationally accessible as retained
battery capacity, we apply a single instantaneous control pulse at an
intermediate time $t_*$,
\begin{align}
U_\theta
=
e^{-{\rm i}\theta\sigma_x}.
\label{eq:RTN_pulse}
\end{align}
At $t_*$, the transverse Bloch component retains the dependence on the
noise history accumulated up to that time. The pulse rotates part of
this component into the longitudinal direction, which is unaffected by
the subsequent pure-dephasing dynamics.

Writing $G(t)=|G(t)|e^{{\rm i}\alpha(t)}$ and
$\vartheta(t)=2\varepsilon t+\phi+\alpha(t)$, the Bloch vector
immediately before the pulse is
\begin{align}
\vec r_-(t_*)
=
|G(t_*)|
\bigl(
\cos\vartheta(t_*),
\sin\vartheta(t_*),
0
\bigr).
\end{align}
After the rotation, the longitudinal component becomes
\begin{align}
z_+(t_*)
=
|G(t_*)|
\sin\vartheta(t_*)
\sin(2\theta).
\label{eq:RTN_z_after_pulse}
\end{align}
Since subsequent pure dephasing suppresses the transverse components
while preserving the population imbalance,
\begin{align}
\vec r_\infty
=
\bigl(0,0,z_+(t_*)\bigr),
\end{align}
and the asymptotically retained capacity is
\begin{align}
\workrange_\infty(t_*,\theta)
=
2\varepsilon |G(t_*)|
\left|
\sin\vartheta(t_*)\sin(2\theta)
\right|.
\label{eq:one_pulse_retention}
\end{align}

Because battery capacity is invariant under unitary conjugation, the
pulse has zero capacity cost under the accounting convention introduced
in Sec.~\ref{subsec:process_tensor}.  Hence, the retained capacity in
Eq.~\eqref{eq:one_pulse_retention} is not attributable to capacity directly supplied
by the controller.

\subsection{Capacity-retention advantage over a matched Markovian reference}
\label{subsec:RTN_advantage}

A nonzero retained capacity alone does not isolate the role of dynamical
memory, since the same single-pulse mechanism can also operate under
Markovian dephasing. To identify a memory-assisted advantage, we
therefore compare the non-Markovian RTN dynamics with a Markovian
reference from the same RTN family, subject to exactly the same class of
single $x$-axis pulses.

For the numerical illustration, we set
$\varepsilon=0.5$ and $\phi=0$, and choose the non-Markovian
RTN parameters
\begin{align}
(\xi_{\rm NM},\nu_{\rm NM})=(1,2),
\end{align}
which lie in the oscillatory regime
$\xi_{\rm NM}<2\nu_{\rm NM}$.
To construct a comparable Markovian reference, we fit its uncontrolled
capacity dynamics to that of the non-Markovian dynamics over the interval
$0\le t\le T_{\rm fit}$ by minimizing
\begin{align}
\mathcal E(\xi_{\rm M},\nu_{\rm M})
=
\int_0^{T_{\rm fit}}
\left[
\workrange_{\rm NM}(\rho(t);H_0)
-
\workrange_{\rm M}(\rho(t);H_0)
\right]^2 dt
\label{eq:SM_fit_M_reference}
\end{align}
subject to the Markovian condition
$\xi_{\rm M}>2\nu_{\rm M}$.
For $T_{\rm fit}=4$, the optimization gives
\begin{align}
(\xi_{\rm M},\nu_{\rm M})
=
(179.12,12.23),
\end{align}
which we refer to as the matched Markovian RTN reference.

Both dynamics are then subjected to the same single-pulse control
family. For $X\in\{{\rm NM},{\rm M}\}$, let
$\rho_X(T;t_*,\theta)$ denote the terminal state obtained by applying
$U_\theta=e^{-{\rm i}\theta\sigma_x}$
at time $t_*$. The optimized retained capacity is defined as
\begin{align}
\workrange_X^{\rm opt}(T)
:=
\max_{\substack{0\le t_*\le T\\ \theta\in[0,\pi)}}
\workrange
\left(
\rho_X(T;t_*,\theta);H_0
\right).
\label{eq:Cx_opt_process}
\end{align}
To retain the system--noise correlations accumulated up to the pulse
time $t_*$, the finite-time controlled evolution is evaluated using the
augmented RTN propagator described in Appendix~\ref{app:one_pulse_optimization}.

For an additional same-control comparison, we also apply the NM-optimized pulse to the Markovian reference. If $(t_*^{\rm NM},\theta^{\rm NM})$ maximizes $\workrange_{\rm NM}^{\rm opt}(T)$, we define
\begin{align}
\workrange_{\rm M}^{*}(T)
:=
\workrange\left(
\rho_{\rm M}(T;t_*^{\rm NM},\theta^{\rm NM});H_0
\right).
\end{align}
By construction,
\begin{align}
\workrange_{\rm M}^{\rm opt}(T)
\geq
\workrange_{\rm M}^{*}(T).
\end{align}
A single-pulse capacity-retention advantage is therefore established when
\begin{align}
\workrange_{\rm NM}^{\rm opt}(T)
>
\workrange_{\rm M}^{\rm opt}(T)
\ge
\workrange_{\rm M}^{*}(T).
\label{eq:NM_one_pulse_advantage}
\end{align}

\begin{figure}[!t]
\centering
\includegraphics[width=1\linewidth]{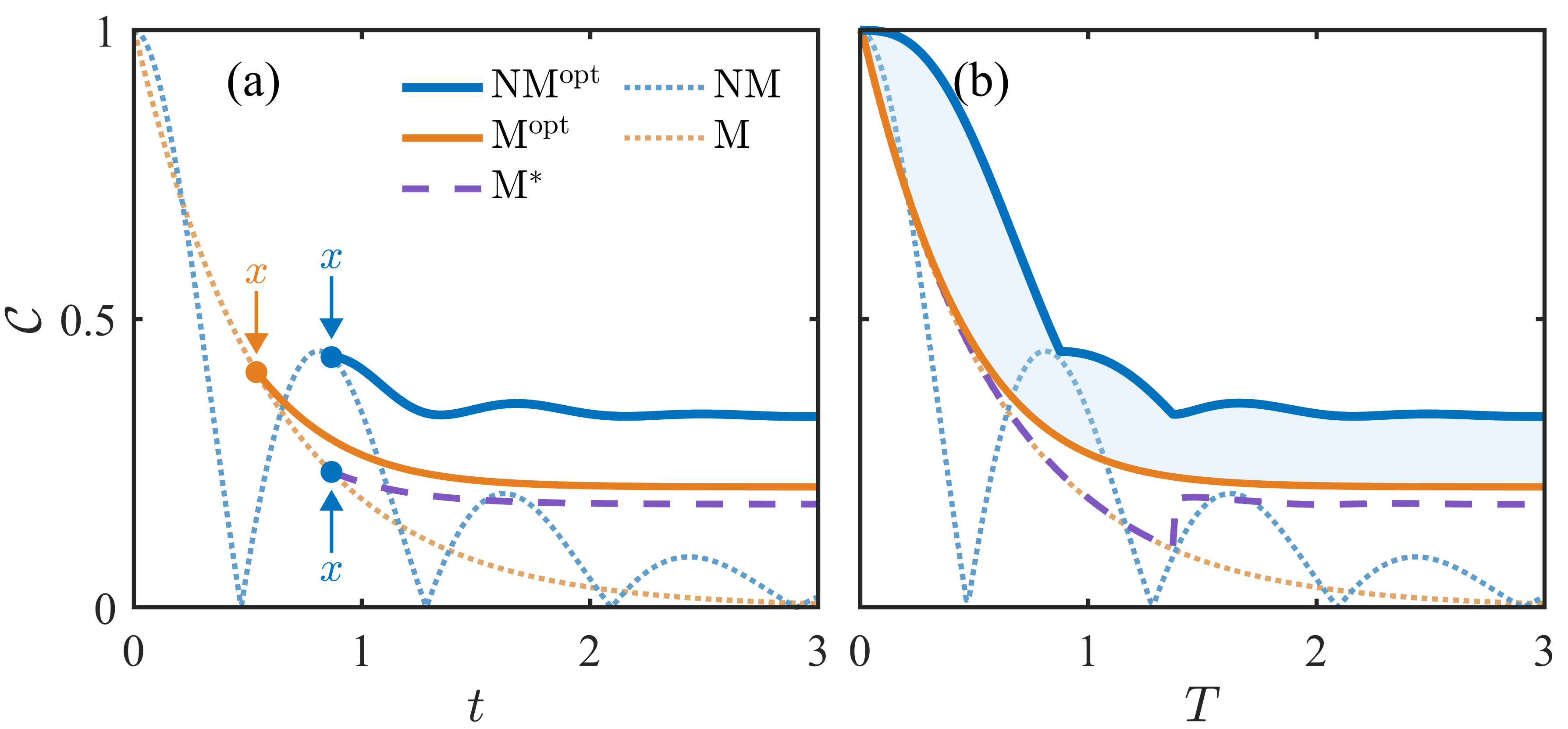}
\caption{
Single-pulse capacity retention under RTN dephasing.
The non-Markovian dynamics has
$(\xi_{\rm NM},\nu_{\rm NM})=(1,2)$, while the matched Markovian
reference has
$(\xi_{\rm M},\nu_{\rm M})=(179.12,12.23)$, obtained with
$T_{\rm fit}=4$.
${\rm NM}^{\rm opt}$ and ${\rm M}^{\rm opt}$ denote independently
optimized single-pulse protocols, and ${\rm M}^{*}$ the Markovian
dynamics driven by the NM-optimized pulse. Dotted curves show the
corresponding free evolutions, and the legend in (a) applies to both
panels.
(a) Controlled trajectories for terminal time $T=3$; dots and arrows
mark the pulse times.
(b) Optimized retained-capacity envelopes obtained by re-optimizing
$(t_*,\theta)$ for each $T$. The shaded region represents
$\mathcal{C}_{\rm NM}^{\rm opt}(T)-\mathcal{C}_{\rm M}^{\rm opt}(T)$.
}
\label{fig:comb1_main}
\end{figure}

Figure~\ref{fig:comb1_main} shows the resulting single-pulse
optimization. For the fixed terminal time $T=3$, the independently
optimized pulses are
\begin{align}
(t_*^{\rm NM},\theta^{\rm NM})
=
(0.867,\pi/4),\quad
(t_*^{\rm M},\theta^{\rm M})
=
(0.538,\pi/4).
\end{align}
As shown in Fig.~\ref{fig:comb1_main}(a), the non-Markovian dynamics
retains a larger terminal capacity than both the independently optimized
Markovian reference and the Markovian trajectory driven by the
NM-optimized pulse. In contrast, the uncontrolled capacity curves decay toward zero. 

Figure~\ref{fig:comb1_main}(b) shows the endpoint-dependent retained capacity obtained by re-optimizing $(t_*,\theta)$ independently for each terminal time $T$.  Over the nontrivial time region shown in the figure, the advantage criterion in Eq.~\eqref{eq:NM_one_pulse_advantage} is satisfied.  The corresponding optimal pulse times and angles are given in Appendix~\ref{app:one_pulse_optimization}. These results demonstrate that memory within the continuous-time non-Markovian RTN evolution can be converted into a measurable capacity-retention advantage.

\section{Discussion}
\label{sec:IV}

Our results identify two complementary settings in which temporal memory can enhance quantum-battery capacity. Within the multitime process-tensor framework, memory-enhanced strategies can outperform locally optimized operations, while the resulting capacity advantage vanishes for Markovian processes and is bounded by a distance-based measure of process non-Markovianity. An explicit three-step construction demonstrates a strict advantage generated by inter-step memory. Complementarily, the RTN proof of principle shows that memory accumulated during continuous evolution can be accessed through an intermediate control operation, yielding a capacity-retention advantage over a matched Markovian reference under the same control class. Together, these results show that temporal memory can serve as an operational resource for quantum-battery capacity rather than merely as a dynamical signature of open-system evolution.

More broadly, these results motivate treating environmental memory as a potential resource to be harnessed through quantum control rather than solely as a source of degradation. An important direction is to determine how the memory-assisted advantages identified here extend to many-body systems and realistic quantum-battery architectures \cite{Ferraro2026,Xu2026}, and how they may be combined with experimentally accessible control and reservoir engineering \cite{Kai2026}. Such extensions may clarify when temporal correlations provide practically useful thermodynamic advantages in quantum energetic devices.

\section*{Acknowledgments}

This work was supported by the National Natural Science Foundation of China (Nos.~62172341, 12204386, and 12075159), Interdisciplinary Research of Southwest Jiaotong University (No.~2682022KJ004), and the Fundamental Research Funds for the Central Universities (No.~2682026CX168).

\section*{Data Availability}
 The data that support the findings of this study are available from the corresponding author upon reasonable request.

\appendix

\section{Proof of Theorem 1}
\label{appendix:Theorem1}

The lower bound follows directly from the strategy hierarchy established
in Sec.~\ref{subsec:process_tensor}. Since the memoryless strategy is contained in the admissible
memory-enhanced strategy class,
\begin{align}
\Delta\workrange^{\rm ME}(\mathcal P;H)
\ge
\Delta\workrange^{\rm ML}(\mathcal P;H),
\label{eq:SM_ME_ge_ML}
\end{align}
and hence
$\mathcal A_{\workrange}(\mathcal P;H)\ge0$.

We next show that the memory-enhanced advantage vanishes for Markovian
process tensors, and then derive the continuity bounds needed to relate
this advantage to the process non-Markovianity.

\begin{lemma}
\label{lem:Markovian_collapse}
For any Markovian $n$-step process tensor $\mathcal P_{\rm M}$,
\begin{align}
\Delta\workrange^{\rm ME}(\mathcal P_{\rm M};H)
=
\Delta\workrange^{\rm ML}(\mathcal P_{\rm M};H).
\label{eq:SM_Markovian_collapse}
\end{align}
\end{lemma}

\begin{proof}
For a Markovian process tensor, the conditional map at each process
slot is independent of the preceding intervention history. Denote the
corresponding channel at step $i$ by $\Lambda_i$. The memoryless
optimization therefore separates into independent single-step
problems,
\begin{align}
\Delta\workrange^{\rm ML}(\mathcal P_{\rm M};H)
=
\sum_{i=1}^{n}
\Delta\workrange(\Lambda_i;H).
\label{eq:SM_ML_Markovian}
\end{align}

Now consider an arbitrary restricted comb strategy
$\mathcal S\in\mathbb S_n^{\rm C}$.
Let $\rho_i$ be the state entering the $i$-th process slot and
\begin{align}
\omega_i=\Lambda_i(\rho_i)
\end{align}
the corresponding output. By definition of the single-step capacity
gain,
\begin{align}
\workrange(\omega_i;H)
-
\workrange(\rho_i;H)
\le
\Delta\workrange(\Lambda_i;H)
\label{eq:SM_slot_gain_bound}
\end{align}
for every $i$.

Let
$\{\mathcal E_0,\mathcal E_1,\ldots,\mathcal E_{n-1}\}$
denote the charged control maps of $\mathcal S$, where
$\mathcal E_0$ prepares the first input. Their total capacity cost is
\begin{align}
\Delta\workrange(\mathcal S;H)
=
\sum_{i=0}^{n-1}
\Delta\workrange(\mathcal E_i;H).
\label{eq:SM_comb_cost_used}
\end{align}
For every control map $\mathcal E_i:S\to S$ and every input state
$\tau$,
\begin{align}
\workrange(\mathcal E_i(\tau);H)
-
\workrange(\tau;H)
\le
\Delta\workrange(\mathcal E_i;H).
\label{eq:SM_control_gain_bound}
\end{align}
Hence no control map can generate a positive capacity contribution that
is not included in the controller cost. In particular, for a replacement
map $\mathcal E_i(\cdot)=\rho$,
\begin{align}
\workrange(\rho;H)
\le
\Delta\workrange(\mathcal E_i;H).
\end{align}

We now combine the process and controller contributions along the causal
chain. Let
\begin{align}
\sigma=\mathcal P_{\rm M}(\mathcal S)
\end{align}
be the final state on the retained output registers
$B_1,\ldots,B_{m(\mathcal S)}$, and let
$\sigma_j={\rm Tr}_{\bar j}\sigma$ denote the marginal on $B_j$.
Whenever the output of the $i$-th process slot is retained as $B_j$,
we have $\sigma_j=\omega_i$.

The initial preparation satisfies
\begin{align}
\workrange(\rho_1;H)
\le
\Delta\workrange(\mathcal E_0;H).
\label{eq:SM_initial_prep_bound}
\end{align}
For an intermediate slot $i<n$, if $\omega_i$ is consumed to prepare
the next input, then Eqs.~\eqref{eq:SM_slot_gain_bound} and
\eqref{eq:SM_control_gain_bound} give
\begin{align}
\workrange(\rho_{i+1};H)
-
\workrange(\rho_i;H)
\le
\Delta\workrange(\Lambda_i;H)
+
\Delta\workrange(\mathcal E_i;H).
\label{eq:SM_consumed_balance}
\end{align}
If instead $\omega_i$ is retained as $B_j$, the next input is prepared
independently by a replacement map, yielding
\begin{align}
&\workrange(\sigma_j;H)
+
\workrange(\rho_{i+1};H)
-
\workrange(\rho_i;H)
\nonumber\\
&\qquad\le
\Delta\workrange(\Lambda_i;H)
+
\Delta\workrange(\mathcal E_i;H).
\label{eq:SM_retained_balance}
\end{align}
For the final slot, the same relation holds without the
$\workrange(\rho_{i+1};H)$ term.

Summing these inequalities along the causal chain cancels all
intermediate input capacities and gives
\begin{align}
\sum_{j=1}^{m(\mathcal S)}
\workrange(\sigma_j;H)
-
\Delta\workrange(\mathcal S;H)
\le
\sum_{i=1}^{n}
\Delta\workrange(\Lambda_i;H).
\label{eq:SM_local_balance}
\end{align}

It remains to evaluate the readout functional. For the Markovian process
$\mathcal P_{\rm M}$, each $\Lambda_i$ acts only on the current input,
while Eq.~\eqref{eq:SM_comb_factorization} ensures that this input is
uncorrelated with all previously retained registers. Since retained
outputs are subsequently passive, it follows inductively that
\begin{align}
\sigma
=
\bigotimes_{j=1}^{m(\mathcal S)}
\sigma_j .
\end{align}
The correlation-adjusted readout therefore reduces to the local value,
\begin{align}
\workrange_{\rm read}
\bigl(
\mathcal P_{\rm M}(\mathcal S);
H_{\rm out}^{\mathcal S}
\bigr)
=
\sum_{j=1}^{m(\mathcal S)}
\workrange(\sigma_j;H).
\label{eq:SM_readout_markovian_product}
\end{align}
Thus no purely collective spectral-rearrangement contribution is counted
as a memory advantage.

Combining Eqs.~\eqref{eq:SM_local_balance} and
\eqref{eq:SM_readout_markovian_product}, and using
Eq.~\eqref{eq:SM_ML_Markovian}, we obtain
\begin{align}
&\workrange_{\rm read}
\bigl(
\mathcal P_{\rm M}(\mathcal S);
H_{\rm out}^{\mathcal S}
\bigr)
-
\Delta\workrange(\mathcal S;H)
\nonumber\\
&\qquad\le
\Delta\workrange^{\rm ML}(\mathcal P_{\rm M};H).
\end{align}
Since this holds for every
$\mathcal S\in\mathbb S_n^{\rm C}$,
maximizing over $\mathcal S$ gives
\begin{align}
\Delta\workrange^{\rm ME}(\mathcal P_{\rm M};H)
\le
\Delta\workrange^{\rm ML}(\mathcal P_{\rm M};H).
\end{align}
Together with Eq.~\eqref{eq:SM_ME_ge_ML}, this proves
Eq.~\eqref{eq:SM_Markovian_collapse}.
\end{proof}

In the definition of the process non-Markovianity measure
$N(\mathcal P)$ in Eq.~\eqref{eq:process_NM_measure}, the supremum over
$\mathbb S_n$ ranges over the full set of compatible $n$-step strategy
combs used to define the generalized comb divergence of
Ref.~\cite{PhysRevA.110.042210}. This optimization domain is distinct
from the restricted charged strategy class $\mathbb S_n^{\rm C}$
employed in the memory-enhanced protocol.

In the finite-dimensional fixed-port setting considered here, the
minimum in Eq.~\eqref{eq:process_NM_measure} is attained. Let
$\mathcal Q^\ast\in\mathbb P_n^{\rm M}$ be a closest Markovian process. Then
\begin{align}
\sup_{\mathcal S\in\mathbb S_n}
\left\|\mathcal P(\mathcal S)-\mathcal Q^\ast(\mathcal S)\right\|_1=2N(\mathcal P).
\label{eq:SM_Q_star_distance}
\end{align}

We next establish the following continuity bound for the battery capacity.

\begin{lemma}
\label{lem:capacity_lipschitz}
For any quantum states $\rho$, $\sigma$ and a fixed Hamiltonian $H$, we have
\begin{align}
\left|\workrange(\rho;H)-\workrange(\sigma;H)\right|\le\Lambda \|\rho-\sigma\|_1,
\label{eq:SM_capacity_lipschitz}
\end{align}
where $\Lambda=\lambda_{\max}(H)-\lambda_{\min}(H)$ is the spectral width of $H$.
\end{lemma}

\begin{proof}
Using the variational representation
\begin{align}
\workrange(\rho;H)=\max_U {\rm Tr}\left(HU\rho U^\dagger\right)-
\min_V {\rm Tr}\left(HV\rho V^\dagger\right),
\end{align} 
and the inequalities $|\max_x f_x-\max_x g_x| \leq \max_x|f_x-g_x|$ and
$|\min_x f_x-\min_x g_x| \leq \max_x|f_x-g_x|$, we obtain
\begin{align}
\left|\workrange(\rho;H)-\workrange(\sigma;H)\right| \leq2\max_U\left|{\rm Tr}(HU\Delta U^\dagger)\right|
\label{eq:SM_capacity_lipschitz_step}
\end{align}
with $\Delta=\rho-\sigma$. Let $c=(\lambda_{\max}(H)+\lambda_{\min}(H))/2$ and $\widetilde H=H-c\mathbbm 1$. Since ${\rm Tr}\Delta=0$ and $\|\widetilde H\|_\infty=\Lambda/2$, using H\"older's inequality \cite{Nielsen_Chuang_2010} and the unitary invariance of the trace norm give
\begin{align}
\left|{\rm Tr}\left(HU\Delta U^\dagger\right)\right|=\left|{\rm Tr}(\widetilde H U\Delta U^\dagger)\right|&\leq\|\widetilde H\|_\infty
\|\Delta\|_1\nonumber\\&=\frac{\Lambda}{2}\|\rho-\sigma\|_1 .
\end{align}
Substituting this bound into Eq.~\eqref{eq:SM_capacity_lipschitz_step} yields
\begin{align}
\left|\workrange(\rho;H)-\workrange(\sigma;H)
\right|
\leq\Lambda \|\rho-\sigma\|_1.
\end{align}
\end{proof}

\begin{lemma}
\label{lem:SM_trace_distance_bounds}
Let $\mathcal Q^\ast\in\mathbb P_n^{\rm M}$ be a closest Markovian
process to $\mathcal P$. Then
\begin{align}
\Delta\workrange^{\rm ME}(\mathcal P;H)
-
\Delta\workrange^{\rm ML}(\mathcal Q^\ast;H)
&\le
2n(n+2)\Lambda N(\mathcal P),
\label{eq:SM_ME_NM_bound}
\\
\Delta\workrange^{\rm ML}(\mathcal Q^\ast;H)
-
\Delta\workrange^{\rm ML}(\mathcal P;H)
&\le
2n\Lambda N(\mathcal P),
\label{eq:SM_ML_Qstar_to_P}
\end{align}
where
$\Lambda=\lambda_{\max}(H)-\lambda_{\min}(H)$.
Consequently,
\begin{align}
\Delta\workrange^{\rm ME}(\mathcal P;H)
-
\Delta\workrange^{\rm ML}(\mathcal P;H)
\le
2n(n+3)\Lambda N(\mathcal P).
\label{eq:SM_intrinsic_ME_NM_bound}
\end{align}
\end{lemma}

\begin{proof}
We first bound the variation of the readout functional under a fixed
restricted-comb strategy. Let
$\mathcal S\in\mathbb S_n^{\rm C}$ retain
$m\le n$ output systems, and define
\begin{align}
\sigma=\mathcal P(\mathcal S),
\qquad
\tau=\mathcal Q^\ast(\mathcal S).
\end{align}
Let
\begin{align}
H_{\rm out}^{\mathcal S}
=
\sum_{j=1}^{m}H_{B_j},
\end{align}
where each local Hamiltonian has spectral width $\Lambda$.
For a retained state $\omega$, write
\begin{align}
L(\omega)
&=
\sum_{j=1}^{m}
\workrange(\omega_j;H_{B_j}),
\\
A(\omega)
&=
\workrange(\omega;H_{\rm out}^{\mathcal S})
-
\workrange\!\left(
\bigotimes_{j=1}^{m}\omega_j;
H_{\rm out}^{\mathcal S}
\right)
+
L(\omega).
\end{align}
By the definition of the correlation-adjusted readout,
\begin{align}
\workrange_{\rm read}
(\omega;H_{\rm out}^{\mathcal S})
=
\max\{L(\omega),A(\omega)\}.
\label{eq:SM_readout_max}
\end{align}

Lemma~\ref{lem:capacity_lipschitz} and contractivity of the trace norm
under partial trace give
\begin{align}
|L(\sigma)-L(\tau)|
\le
m\Lambda\|\sigma-\tau\|_1.
\label{eq:SM_L_lip}
\end{align}
Since the spectral width of
$H_{\rm out}^{\mathcal S}$ is at most $m\Lambda$,
\begin{align}
\left|
\workrange(\sigma;H_{\rm out}^{\mathcal S})
-
\workrange(\tau;H_{\rm out}^{\mathcal S})
\right|
\le
m\Lambda\|\sigma-\tau\|_1.
\label{eq:SM_global_lip}
\end{align}
Moreover,
\begin{align}
\left\|
\bigotimes_{j=1}^{m}\sigma_j
-
\bigotimes_{j=1}^{m}\tau_j
\right\|_1
&\le
\sum_{j=1}^{m}
\|\sigma_j-\tau_j\|_1
\nonumber\\
&\le
m\|\sigma-\tau\|_1,
\end{align}
so another application of
Lemma~\ref{lem:capacity_lipschitz} yields
\begin{align}
&
\left|
\workrange\!\left(
\bigotimes_{j=1}^{m}\sigma_j;
H_{\rm out}^{\mathcal S}
\right)
-
\workrange\!\left(
\bigotimes_{j=1}^{m}\tau_j;
H_{\rm out}^{\mathcal S}
\right)
\right|
\nonumber\\
&\qquad\le
m^2\Lambda\|\sigma-\tau\|_1.
\label{eq:SM_product_lip}
\end{align}
Combining these estimates gives
\begin{align}
|A(\sigma)-A(\tau)|
\le
m(m+2)\Lambda\|\sigma-\tau\|_1.
\label{eq:SM_A_lip}
\end{align}
Using
\begin{align}
|\max\{a,b\}-\max\{c,d\}|
\le
\max\{|a-c|,|b-d|\},
\end{align}
Eqs.~\eqref{eq:SM_L_lip} and
\eqref{eq:SM_A_lip} imply
\begin{align}
&
\left|
\workrange_{\rm read}
(\sigma;H_{\rm out}^{\mathcal S})
-
\workrange_{\rm read}
(\tau;H_{\rm out}^{\mathcal S})
\right|
\nonumber\\
&\qquad\le
m(m+2)\Lambda\|\sigma-\tau\|_1.
\label{eq:SM_Cread_continuity}
\end{align}

We now compare the memory-enhanced gains of
$\mathcal P$ and $\mathcal Q^\ast$. Let
$\mathcal S_\varepsilon\in\mathbb S_n^{\rm C}$ be an
$\varepsilon$-optimal strategy for $\mathcal P$, so that
\begin{align}
\Delta\workrange^{\rm ME}(\mathcal P;H)
\le
F_{\mathcal P}(\mathcal S_\varepsilon)+\varepsilon,
\end{align}
where
\begin{align}
F_{\mathcal R}(\mathcal S)
:=
\workrange_{\rm read}
\bigl(\mathcal R(\mathcal S);H_{\rm out}^{\mathcal S}\bigr)
-
\Delta\workrange(\mathcal S;H).
\end{align}
Since the same strategy is admissible for $\mathcal Q^\ast$,
\begin{align}
\Delta\workrange^{\rm ME}(\mathcal Q^\ast;H)
\ge
F_{\mathcal Q^\ast}(\mathcal S_\varepsilon).
\end{align}
Therefore,
\begin{align}
&\Delta\workrange^{\rm ME}(\mathcal P;H)
-
\Delta\workrange^{\rm ME}(\mathcal Q^\ast;H)
\nonumber\\
&\le
F_{\mathcal P}(\mathcal S_\varepsilon)
-
F_{\mathcal Q^\ast}(\mathcal S_\varepsilon)
+\varepsilon.
\end{align}
For the same strategy, the controller cost is identical for the two
processes and therefore cancels. Hence, using
Eq.~\eqref{eq:SM_Cread_continuity},
\begin{align}
&\Delta\workrange^{\rm ME}(\mathcal P;H)
-
\Delta\workrange^{\rm ME}(\mathcal Q^\ast;H)
\nonumber\\
&\le
m(m+2)\Lambda
\left\|
\mathcal P(\mathcal S_\varepsilon)
-
\mathcal Q^\ast(\mathcal S_\varepsilon)
\right\|_1
+\varepsilon.
\end{align}
Since $\mathbb S_n^{\rm C}\subseteq\mathbb S_n$, the definition of
$N(\mathcal P)$ gives
\begin{align}
\left\|
\mathcal P(\mathcal S_\varepsilon)
-
\mathcal Q^\ast(\mathcal S_\varepsilon)
\right\|_1
\le
2N(\mathcal P).
\end{align}
Using $m\le n$ and taking $\varepsilon\to0$ yields
\begin{align}
\Delta\workrange^{\rm ME}(\mathcal P;H)
-
\Delta\workrange^{\rm ME}(\mathcal Q^\ast;H)
\le
2n(n+2)\Lambda N(\mathcal P).
\end{align}
Using the Markovian collapse,
\begin{align}
\Delta\workrange^{\rm ME}(\mathcal Q^\ast;H)
=
\Delta\workrange^{\rm ML}(\mathcal Q^\ast;H),
\end{align}
proves Eq.~\eqref{eq:SM_ME_NM_bound}.

It remains to compare the memoryless benchmarks of
$\mathcal P$ and $\mathcal Q^\ast$.
Let
\begin{align}
\bm r^{\rm ML}
=
(\rho_1^{\rm ML},\ldots,\rho_n^{\rm ML})
\end{align}
be the memoryless trajectory generated for $\mathcal P$, and write
\begin{align}
\Phi_i
:=
\mathcal P_{i|\bm r^{\rm ML}_{<i}} .
\end{align}
Because $\mathcal Q^\ast$ is Markovian, its $i$-th conditional channel
is independent of the preceding trajectory; denote it by $\Lambda_i$.
Let $\rho_i$ be an optimizing input for
$\Delta\workrange(\Lambda_i;H)$. Then
\begin{align}
\Delta\workrange(\Lambda_i;H)
-
\Delta\workrange(\Phi_i;H)
&\le
\workrange(\Lambda_i(\rho_i);H)
-
\workrange(\Phi_i(\rho_i);H)
\nonumber\\
&\le
\Lambda
\left\|
\Lambda_i(\rho_i)
-
\Phi_i(\rho_i)
\right\|_1 .
\label{eq:SM_single_ML_compare}
\end{align}
The two states on the right-hand side can be realized as corresponding
marginals of the outputs of the same compatible strategy applied to
$\mathcal Q^\ast$ and $\mathcal P$, respectively. Hence trace-norm
contractivity and the definition of $N(\mathcal P)$ imply
\begin{align}
\left\|
\Lambda_i(\rho_i)
-
\Phi_i(\rho_i)
\right\|_1
\le
2N(\mathcal P).
\end{align}
Therefore,
\begin{align}
\Delta\workrange(\Lambda_i;H)
-
\Delta\workrange(\Phi_i;H)
\le
2\Lambda N(\mathcal P).
\end{align}
Summing over the $n$ steps and using
\begin{align}
\Delta\workrange^{\rm ML}(\mathcal Q^\ast;H)
&=
\sum_{i=1}^{n}
\Delta\workrange(\Lambda_i;H),
\\
\Delta\workrange^{\rm ML}(\mathcal P;H)
&=
\sum_{i=1}^{n}
\Delta\workrange(\Phi_i;H),
\end{align}
gives Eq.~\eqref{eq:SM_ML_Qstar_to_P}.

Finally, combining
Eqs.~\eqref{eq:SM_ME_NM_bound} and
\eqref{eq:SM_ML_Qstar_to_P}, we obtain
\begin{align}
&
\Delta\workrange^{\rm ME}(\mathcal P;H)
-
\Delta\workrange^{\rm ML}(\mathcal P;H)
\nonumber\\
&=
\Delta\workrange^{\rm ME}(\mathcal P;H)
-
\Delta\workrange^{\rm ML}(\mathcal Q^\ast;H)
\nonumber\\
&\quad+
\Delta\workrange^{\rm ML}(\mathcal Q^\ast;H)
-
\Delta\workrange^{\rm ML}(\mathcal P;H)
\nonumber\\
&\le
2n(n+3)\Lambda N(\mathcal P),
\end{align}
which proves Eq.~\eqref{eq:SM_intrinsic_ME_NM_bound}.
\end{proof}

\section{Finite-time propagation and one-pulse optimization for RTN dynamics}
\label{app:one_pulse_optimization}

This appendix gives the finite-time propagation used for the one-pulse
optimization in Sec.~\ref{subsec:RTN_advantage}. For uninterrupted
free evolution, the reduced dynamics is fully characterized by the
dephasing function $G(t)$. After an intermediate pulse at time $t_*$,
however, the subsequent evolution cannot in general be obtained by
propagating the reduced state $\rho(t_*)$ alone. The RTN is temporally
correlated, so the system state at $t_*$ retains statistical correlations
with the instantaneous noise variable, which affect the evolution after
the pulse. We therefore propagate the system Bloch vector together with
a noise-weighted moment that retains this information.

For a fixed RTN realization $c$, write
\begin{align}
\rho_c(t)
=
\frac12
\left[
\mathbbm 1+\vec r_c(t)\cdot\vec\sigma
\right].
\end{align}
We use the averaged Bloch vector together with the noise-weighted
moment,
\begin{align}
\bm v(t)=
\begin{pmatrix}
\vec r(t)\\
\vec\eta(t)
\end{pmatrix}
=
(x(t),y(t),z(t),p(t),q(t),w(t))^T,
\end{align}
where $\vec r(t)=\langle \vec r_c(t)\rangle_c$ and $\vec\eta(t)=\langle c(t)\vec r_c(t)\rangle_c$. Only the first three components are physical Bloch components. For $H_0=\varepsilon\sigma_z$, the capacity is therefore given by 
\begin{align}
\workrange(t)=2\varepsilon
\sqrt{x^2(t)+y^2(t)+z^2(t)}
=2\varepsilon \sqrt{\bm v^T(t)M\bm v(t)},
\label{eq:SM_capacity_augmented}
\end{align}
with $M=\mathrm{diag}(1,1,1,0,0,0)$.

A free RTN segment of duration $u$ is propagated exactly as \cite{PhysRevA.78.022313}
\begin{align}
\bm v(t+u)
=
F_{\xi,\nu}(u)\bm v(t).
\end{align}
For $\xi<2\nu$, let
\begin{align}
\omega&=\sqrt{4\nu^2-\xi^2},
\\
a(u)&=
e^{-\xi u}
\left[
\cos(\omega u)
+\frac{\xi}{\omega}\sin(\omega u)
\right],
\\
\beta(u)&=
\frac{2\nu}{\omega}
e^{-\xi u}\sin(\omega u),
\\
h(u)&=
e^{-\xi u}
\left[
\cos(\omega u)
-\frac{\xi}{\omega}\sin(\omega u)
\right],
\\
d(u)&=e^{-2\xi u}.
\end{align}
For $\xi>2\nu$, with
$\Omega=\sqrt{\xi^2-4\nu^2}$, the corresponding expressions are
\begin{align}
a(u)&=
e^{-\xi u}
\left[
\cosh(\Omega u)
+\frac{\xi}{\Omega}\sinh(\Omega u)
\right],
\\
\beta(u)&=
\frac{2\nu}{\Omega}
e^{-\xi u}\sinh(\Omega u),
\\
h(u)&=
e^{-\xi u}
\left[
\cosh(\Omega u)
-\frac{\xi}{\Omega}\sinh(\Omega u)
\right],
\end{align}
with $d(u)=e^{-2\xi u}$ unchanged.

Writing
\begin{align}
C_u=\cos(2\varepsilon u),
\qquad
S_u=\sin(2\varepsilon u),
\end{align}
the augmented propagator is
\begin{align}
F_{\xi,\nu}(u)=
\begin{pmatrix}
a C_u & -a S_u & 0 & -\beta S_u & -\beta C_u & 0\\
a S_u &  a C_u & 0 &  \beta C_u & -\beta S_u & 0\\
0     &  0     & 1 &  0         &  0         & 0\\
-\beta S_u & -\beta C_u & 0 & h C_u & -h S_u & 0\\
\beta C_u  & -\beta S_u & 0 & h S_u &  h C_u & 0\\
0          & 0          & 0 & 0     &  0     & d
\end{pmatrix},
\label{eq:SM_F_RTN}
\end{align}
where $a,\beta,h$, and $d$ are evaluated at $u$. This is the exact
ensemble-averaged RTN propagator in the augmented representation.

For the equatorial initial state used in the main text,
\begin{align}
\bm v_0
=
(\cos\phi,\sin\phi,0,0,0,0)^T.
\end{align}
Immediately before the pulse,
\begin{align}
\bm v_-(t_*)
=
F_{\xi,\nu}(t_*)\bm v_0.
\label{eq:SM_v_before_pulse}
\end{align}
The additional components
$\vec\eta_-(t_*)=\langle c(t_*)\vec r_c(t_*)\rangle_c$
retain the system--noise correlations required to propagate the
post-pulse dynamics.

The pulse
$U_\theta=e^{{\rm -i}\theta\sigma_x}$ rotates both
$\vec r$ and $\vec\eta$ by $R_x(2\theta)$, so that
\begin{align}
\bm v_+(t_*)
=
P(\theta)\bm v_-(t_*),
\qquad
P(\theta)
=
R_x(2\theta)\oplus R_x(2\theta),
\label{eq:SM_P_pulse}
\end{align}
with
\begin{align}
R_x(2\theta)
=
\begin{pmatrix}
1 & 0 & 0\\
0 & \cos(2\theta) & -\sin(2\theta)\\
0 & \sin(2\theta) & \cos(2\theta)
\end{pmatrix}.
\end{align}
The final augmented state is therefore
\begin{align}
\bm v_T^{(\xi,\nu)}(t_*,\theta)
=
F_{\xi,\nu}(T-t_*)
P(\theta)
F_{\xi,\nu}(t_*)
\bm v_0 .
\label{eq:SM_vT_onepulse}
\end{align}
Its terminal capacity is
\begin{align}
\workrange_T^{(\xi,\nu)}(t_*,\theta)
=
2\varepsilon
\sqrt{
\left[
\bm v_T^{(\xi,\nu)}(t_*,\theta)
\right]^T
M
\bm v_T^{(\xi,\nu)}(t_*,\theta)
}.
\label{eq:SM_CT_onepulse}
\end{align}

For each terminal time $T$, Fig.~\ref{fig:comb1_main}(b) is obtained by
independently maximizing
\begin{align}
\workrange_{(\xi,\nu)}^{\rm opt}(T)
=
\max_{\substack{0\le t_*\le T\\ \theta\in[0,\pi)}}
\workrange_T^{(\xi,\nu)}(t_*,\theta).
\label{eq:SM_onepulse_optimization}
\end{align}

\begin{figure}[ht!]
\centering
\includegraphics[width=\linewidth]{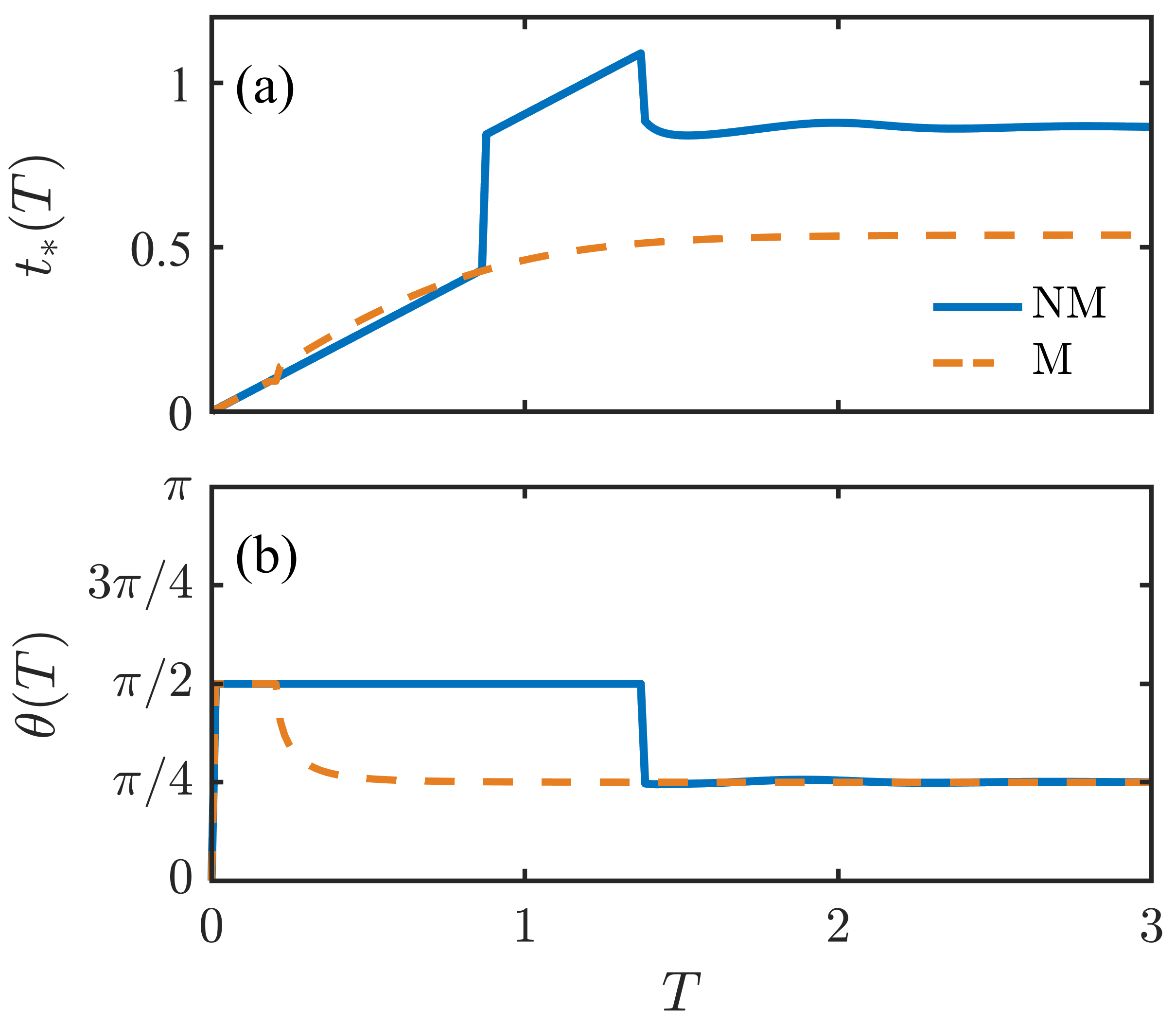}
\caption{
Optimal pulse parameters corresponding to the retained-capacity
envelopes in Fig.~\ref{fig:comb1_main}(b).
The optimization is performed independently for each terminal time $T$
for the non-Markovian RTN dynamics and the matched Markovian reference.
(a) Optimal pulse time $t_*(T)$.
(b) Optimal pulse angle $\theta(T)$.
Jumps indicate switches between competing optimum branches.
}
\label{fig:opt_pulse_parameters}
\end{figure}

Figure~\ref{fig:opt_pulse_parameters} shows the pulse parameters
associated with the optimized envelopes. The jumps arise from switches
between competing optimum branches and do not imply discontinuities of
the optimized capacity. At longer times, the selected optimal branches approach
$\theta=\pi/4$, one of the maximizing values of
$|\sin(2\theta)|$ in Eq.~\eqref{eq:one_pulse_retention}.

%


\end{document}